\documentclass{article}
\usepackage{authblk}
\usepackage{graphicx}
\usepackage[latin1]{inputenc}
\usepackage{epsfig}
\usepackage{amssymb}
\usepackage[english]{babel}
\usepackage{setspace}
\usepackage{graphicx}
\usepackage{amssymb}
\usepackage[perpage,symbol*]{footmisc}
\usepackage{amsfonts}
\usepackage{comment}
\usepackage{graphics}
\usepackage{amsmath}
\begin{document}
\title{Woods-Saxon equivalent to a double folding potential}
\author{ A. S. Freitas, L. Marques, X. X. Zhang, M. A. Luzio, P. Guillaumon, R. Pampa Condori, R.~Lichtenth\"aler}
\affil{Instituto de F\'{i}sica da Universidade de S\~ao Paulo, C.P. 66318, 05389-970, S\~ao Paulo, Brazil.}
\maketitle

\begin{abstract}

A Woods-Saxon equivalent to a double folding potential in the surface region is obtained for the heavy-ion scattering potential.
The Woods-Saxon potential has fixed geometry and was used as a bare potential in the analysis of elastic scattering angular distributions 
of several stable systems.
A new analytical formula for the position and height of the Coulomb barrier is presented, which reproduces the results obtained
using double folding potentials. This simple formula has been applied to estimate the fusion cross section above the Coulomb barrier.
A comparison with experimental data is presented. 

\end{abstract}


\section{Introduction}

The strong nuclear force is responsible for keeping the nucleons together inside the nucleus and is still not fully understood.
The attractive force between the nucleons is the residuum of the interaction  between quarks and gluons confined inside the nucleons and the connection
between  the fundamental interaction and the nucleon-nucleon force is still an open problem. Elastic scattering between nuclei provides information of the nuclear interaction
however, the cross sections are 
affected by the couplings between the elastic scattering and all other possible reaction channels. 
The potential obtained from the analysis of elastic scattering angular distributions is the sum of a bare potential, which is real in principle, and a polarization term, 
which is complex and contains the effects of all couplings. To obtain information of the bare potential one should find a situation where the elastic is the only open channel,
however, it is very difficult to find such experimental situation. 
Even at energies around the Coulomb barrier, where the reactions channels are closing, there is still the contribution of the fusion process, which makes the interacting 
potential complex.
If we go down to even lower energies, the effect of the short range nuclear potential becomes smaller and smaller as the long range Coulomb potential
dominates, making the scattering pure Rutherford. 

The bare potential can be in principle be defined as the result of the double folding of the nucleon-nucleon interactions and  the projectile and target nuclear 
densities\cite{satchler,satchler1,love,bertsch}. The nucleon-nucleon interactions can be obtained from more fundamental theories. 
Double folding potentials have been used as the bare potential to analyse 
experimental data \cite{mohr} and the imaginary part of the interaction is normally parameterized and freely searched to best fit the angular distributions.  
One of the most widely used parameterizations for the nuclear potential is the well known 
Woods-Saxon (WS) shape \cite{woods-saxon}, with three parameters that are adjusted to reproduce the data.

More recently, the S\~ao Paulo optical potential (PSP) has been developed, where the real part is taken as a double folding potential and
the imaginary part has the same geometry of the real part, with an additional fixed normalization factor.
An energy dependence term is included to account for non-local corrections due to the Pauli principle \cite{chamon1,chamon}.
The S\~ao Paulo potential has no free parameters and has been succesfully applied to a large number of experimental angular distributions 
from low to intermediate energies.

Despite the success of the S\~ao Paulo potential in analysing elastic scattering data, it would be interesting to investigate
the relation between the Woods-Saxon shape and the double folding potential.
Most of the optical model and reaction programs
use the WS parameterization whereas double folding potentials have to be entered externally from numerical
files. The equivalence between WS and double folding potentials is not straightforward and there is no WS that could reproduce
the double folding shape in the whole radial range. However, heavy ion  scattering at low energies
is frequently sensitive only to the tail of the nuclear potential and not very much to the potential in the interior region,
where strong absorption usually takes place.

In the present work we determine the parameters of a WS potential that reproduce the real part of the double folding
S\~ao Paulo potential in the surface region.
We determine the geometry of this potential and apply it to experimental elastic scattering angular distributions
of several systems. 

Based on the real part of this potential we found a simple analytical formula to obtain the position and the height 
of the Coulomb barrier which reproduces quite precisely the Coulomb barriers obtained from the S\~ao Paulo potential. 
This formulation is applied to estimate the fusion cross section in the region above the Coulomb barrier.

\section{The Woods-Saxon potential}

The Woods-Saxon shape is given by:

\begin{equation}
f(r)=\frac{1}{1+\exp(\frac{r-R}{a})} \label{eq1}
\end{equation}
where  $R=r_0(A_1^{1/3}+A_2^{1/3})$ and $a$ is the difuseness. The optical potential is written as:

\begin{equation}
V_{nucl}(r)=-V_0f_1(r)-iW_0f_2(r) \label{eq2}
\end{equation}
where $V_0$ and $W_0$ are the real and imaginary strengths of the optical potential
respectively and $f_i(r)$ are Woods-Saxon form factors that may have different values of radius and diffuseness parameters.

We developed a simple computer program to perform an automatic search on the 3 parameters $V_0$, $r_{0}$, $a$ of Eq. \ref{eq1}  to reproduce the tail of  
the real part of the  S\~ao Paulo potential for several systems.  

\begin{figure}
\begin{center}
\includegraphics[width=18pc]{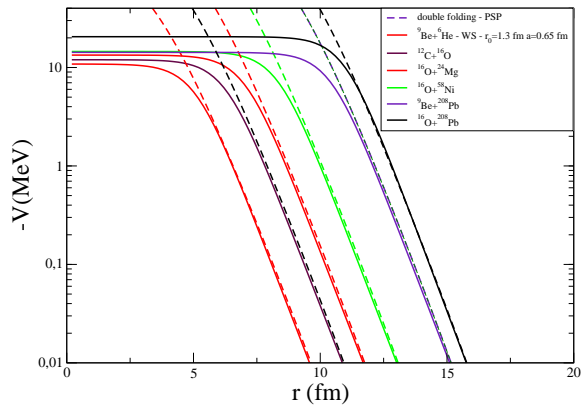}
\end{center}
\caption{\label{fig1} Radial dependence of the real part of the nuclear potential 
for several systems.}
\end{figure}

The results are shown in Fig. \ref{fig1} where the dashed lines represent the real part of S\~ao Paulo potential (PSP) calculated for several systems. The solid lines are the 
resulting Woods-Saxon potentials that best fit the PSP in the region $r>R$.  
We found that, for all systems analysed, the Woods-Saxon that best reproduces the tail of the PSP potential has a 
diffuseness $a \approx0.65$ fm, and $r_0 \approx 1.3$ fm  for strengths ranging between $V_0=10-20$ MeV.  
In general, double folding potentials are strongly attractive with strengths of hundreds of MeV's.

\subsection{Ambiguities}

Ambiguities in the optical potential have been  the subject of many studies since the beginnings of nuclear physics \cite{igo}. The fact that the elastic
scattering angular distributions in the strong absorption regime are sensitive to a small region  in the surface of the nuclear potential, 
has long been recognized. If the scattering at low energies is sensitive only to the tail of the nuclear potential, for a Woods-Saxon shape one immediately gets that,
for $(r-R)>>a$, we obtain  $V_{nucl}(r)=-V_0\exp(R/a)\exp(-r/a)$. Thus, for a given difuseness $a$, any combination of $V_0$ and $R$ that leaves $V_0\exp(R/a)$ unchanged will provide the
same potential in the surface region. In this sense the parameters $V_0$ and $r_0$ proposed in the present paper are only one family of possible potentials.

\section{The Coulomb barrier}

The height and position of the Coulomb barrier are very important parameters in the collision of two heavy ions. They basically determine the total reaction 
cross section at energies above the Coulomb barrier and can be obtained if the real part of the nuclear potential in the 
surface region is known. 

The condition:
\begin{equation}
\frac{d}{dr}[V_{nucl}(r)+V_{coul}(r)]_{r=R_B}=0 \label{eq::RB}
\end{equation}
determines the position $R_B$, and the height of the Coulomb barrier is given by $V_B=V_{nucl}(R_B)+V_{coul}(R_B)$.
As usually $V_{nucl}$ is unknown, we may assume an approximate radius for the Coulomb barrier radius as:

\begin{equation}
\mathring{R}_B=R=r_0(A_1^{1/3}+A_2^{1/3}) \label{eq::cw}
\end{equation}

and use a simplified formula for the height of the Coulomb barrier:

\begin{equation}
 \mathring{V}_{B}=\frac{Z_1Z_2e^2}{\mathring{R}_B}  \label{eq::rb0}
\end{equation}

In general  Eqs. \ref{eq::cw}  and \ref{eq::rb0} do not yield very good results in comparison to  Coulomb barriers obtained from realistic DF potentials.
Eq. \ref{eq::RB} provides a position
for the Coulomb barrier which is, in most cases, larger than the geometrical radius from Eq.\ref{eq::cw} provides similar values of $R_B$.
Indeed, we see that, if we take a Woods-Saxon form for the nuclear potential $V_{nucl}(r)=-V_0/[1+\exp[(r-R)/a]]$ and $V_{coul}(r)=Z_1Z_2e^2/r$ in Eq. \ref{eq::RB} one shows that,
in the approximation $\exp((R_B-R)/a)>>1$, the Coulomb barrier radius can be written as:

\begin{equation}
R_B=R+a \ln\left[\frac{R}{a}\times\frac{V_0}{\mathring{V}_B}\right] \label{eq::rb}
\end{equation}

This equation may not be exact but it displays the main physics of the relation between $R$ and $R_B$. 
For a square potential ($a=0$), we get  $R_B=R$ and there
is no correction to $R$. As the difuseness of the potential increases, the correction term in right hand side of Eq. \ref{eq::rb} increases. Also, for larger Coulomb barriers,
the correction decreases.

Taking $R=1.3(A_1^{1/3}+A_2^{1/3})$ fm,  $a=0.65$ fm, $V_0=15$ MeV and $\mathring{V}_B$ given by Eq. \ref{eq::rb0} one obtains:

\begin{equation}
R_B=R+0.65 \ln[x] \label{eq::rb1}
\end{equation}

where $x=27.1\times\frac{(A_1^{1/3}+A_2^{1/3})^2}{Z_1Z_2}$ is a positive dimensionless parameter:

Then we get for $V_B$:
\begin{equation}
V_B=\frac{Z_1Z_2e^2}{R_B}-\frac{15}{x+1} \label{vcb}
\end{equation}

Equations \ref{eq::rb1} and \ref{vcb} depend only on the masses and charges of the nuclei and provide the Coulomb barrier position and height in very good agreement with those 
obtained from 
numerical calculations using the S\~ao Paulo Potential. This is shown in Table \ref{tab0}. 
The discrepancies are in the order of a few percent or smaller than that for heavier systems. 

\subsection{The curvature of the Coulomb barrier}

One could go a step further and obtain the curvature of the Coulomb barrier based on the above potential. The region around the top of the Coulomb barrier can be 
approximated by an inverted harmonic oscillator potential of height $V_B$ and frequency $w$. The frequency is related to $V_B$ by:
\begin{equation}
\hbar w=\hbar\sqrt{(|d^2V(r)/dr^2|)_{r=R_B}/\mu} \label{hw}
\end{equation}
where $V(r)=V_{nucl}(r)+V_{coul}(r)$, $\mu$ is the reduced mass and:

\begin{equation}
(d^2V(r)/dr^2)_{r=R_B}=-\frac{V_0}{a^2}\frac{x(x-1)}{(x+1)^3}+\frac{2Z_1Z_2e^2}{R_B^3} \label{2cnd}
\end{equation}
Substituting $V_0=15$ MeV and $a=0.65$ fm in the above formula, one can estimate the curvature of the Coulomb barrier.

\begin{table}[h]
\caption{\label{tab0}Comparison between formulas \ref{eq::rb1} and \ref{vcb} (WS) and the results for Coulomb barrier radius and height 
from numerical calculations using the double folding S\~ao Paulo potential (PSP).}
\begin{center}
\begin{tabular}{c||c|c|c|c|c}
\multicolumn{1}{c}{}   & \multicolumn{1}{c}{} & \multicolumn{2}{c}{$R_B$(fm)} &  \multicolumn{2}{c}{$V_B$(MeV)} \\
\cline{3-6}
System &   x   &  \multicolumn{1}{c}{WS}  &   \multicolumn{1}{c}{PSP} &  \multicolumn{1}{|c}{WS}  &   \multicolumn{1}{c}{PSP} \\
\hline
$^6$He$+^9$Be       & 51.45	&  7.62  & 8.00  &	1.22  &	1.32 \\
$^{12}$C$+^{16}$O   & 13.06	&  7.92  & 8.15  &	7.65  &	7.78 \\
$^{16}$O$+^{24}$Mg  & 8.24	&  8.39  & 8.55  &	14.84 &	14.85 \\
$^{16}$O$+^{58}$Ni  & 4.94	&  9.34  & 9.40  &	31.99 &	31.68 \\
$^9$Be$+^{208}$Pb   & 5.29	&  11.48 & 11.50 &	38.72 &	38.55 \\
$^{16}$O$+^{208}$Pb & 2.94	&  11.68 & 11.65 &	77.07 &	75.90 \\
\end{tabular}
\end{center}
\end{table}

\section{Analysis of experimental data}

In the next sections we use this potential to analyse a number of experimental angular distributions. We fix the geometry of the real and imaginary parts to
 $r_0=1.3$ fm and $a=0.65$ fm and allow $V_0$ and $W_0$  vary to best fit the angular distributions. The idea here is not to obtain excelent fits with such a simple 
potential but to show that it is possible to reproduce the general features of the angular distributions using this fixed geometry potential.

All the Optical Model calculations have been performed
using the program SFRESCO, the automatic search version of the FRESCO program \cite{ian}. 
The results are presented in the next subsections.

\subsection{$^{9}$Be$+^{27\!}$Al system} \label{sec-27al}

\begin{figure}
\begin{center}
\includegraphics[width=20pc]{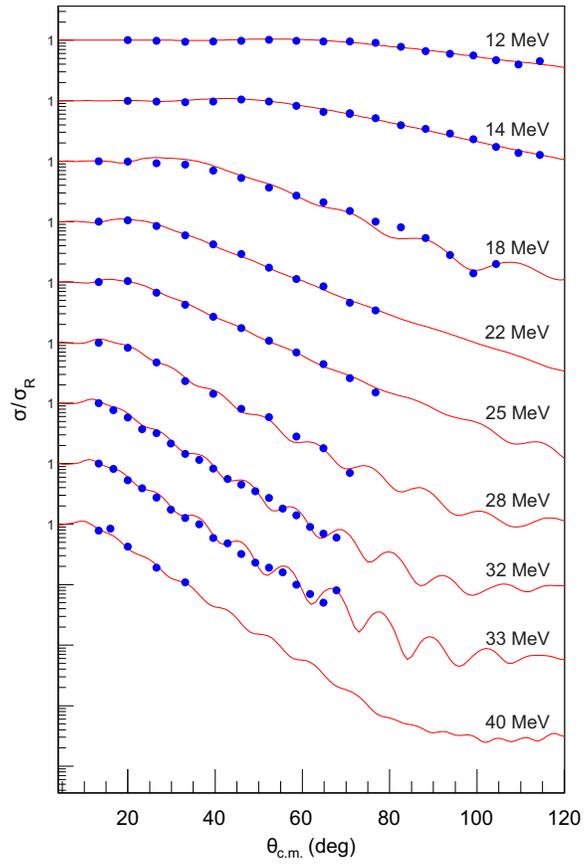}
\end{center}
\caption{\label{fig2} Cross section for the ${^9}$Be+$^{27\!}$Al system.  The energies values are shown in table \ref{tab1}. The experimental data is from Ref. \cite{gomes}. }
\end{figure}

Nine elastic scattering $^{9}$Be+$^{27\!}$Al angular distributions measured by  P. R. S. Gomes et al. \cite{gomes0} have been analysed in the range from
$12$ MeV to $40$ MeV in the laboratory system. Equation \ref{vcb} gives $V_{B}=10.65$ MeV for the Coulomb barrier energy  in 
the laboratory system. The geometry of the real and imaginary potentials was fixed at $r_o=1.3$ fm and $a=0.65$ fm and the depths $V_0$
and $W_0$ were  varied to best fit the data. The Coulomb radius parameter was fixed at $r_{0c}=1.3$ fm.
The results are shown in Figure \ref{fig2} and the fitted parameters are presented in Table \ref{tab1} together with the errors and the best reduced chi-square
values. The errors have been estimated by the program SFRESCO using the gradient method used in the search procedure.

\begin{table}[h]
\caption{\label{tab1}Parameters of WS potential for the system $^{9}$Be+$^{27\!}$Al.}
\begin{center}
\begin{tabular}{c|c|c|c}
$E_{lab}$  (MeV) & $V_0$ (MeV) & $W_0$ (MeV) & $\chi^2$ \\
12.0 & 	18.15$\pm$	2.10 &	15.30$\pm$	7.68 &	0.36 \\
14.0 & 	15.22$\pm$ 	1.46 &	10.08$\pm$	2.90 &	0.47 \\	
18.0 & 	17.87$\pm$	0.21 &	3.37$\pm$	0.17 &	4.35 \\
22.0 & 	16.45$\pm$	2.72 &	16.50$\pm$	4.05 &	0.69 \\
25.0 & 	12.22$\pm$	0.30 &	13.64$\pm$	0.01 &	1.83 \\
28.0 & 	17.39$\pm$	0.75 &	12.17$\pm$	0.52 &	3.92 \\
32.0 & 	15.90$\pm$	0.62 &	11.96$\pm$	0.52 &	6.24 \\
33.0 & 	17.43$\pm$	0.48 &	12.12$\pm$	0.35 &	13.36 \\
40.0 & 	8.29$\pm$	1.33 &	10.62$\pm$	1.85 &	2.70 \\
\end{tabular}
\end{center}
\end{table}

\subsection{$^{16}$O+$^{58}$Ni system}

\begin{figure}
\begin{center}
\includegraphics[width=18pc]{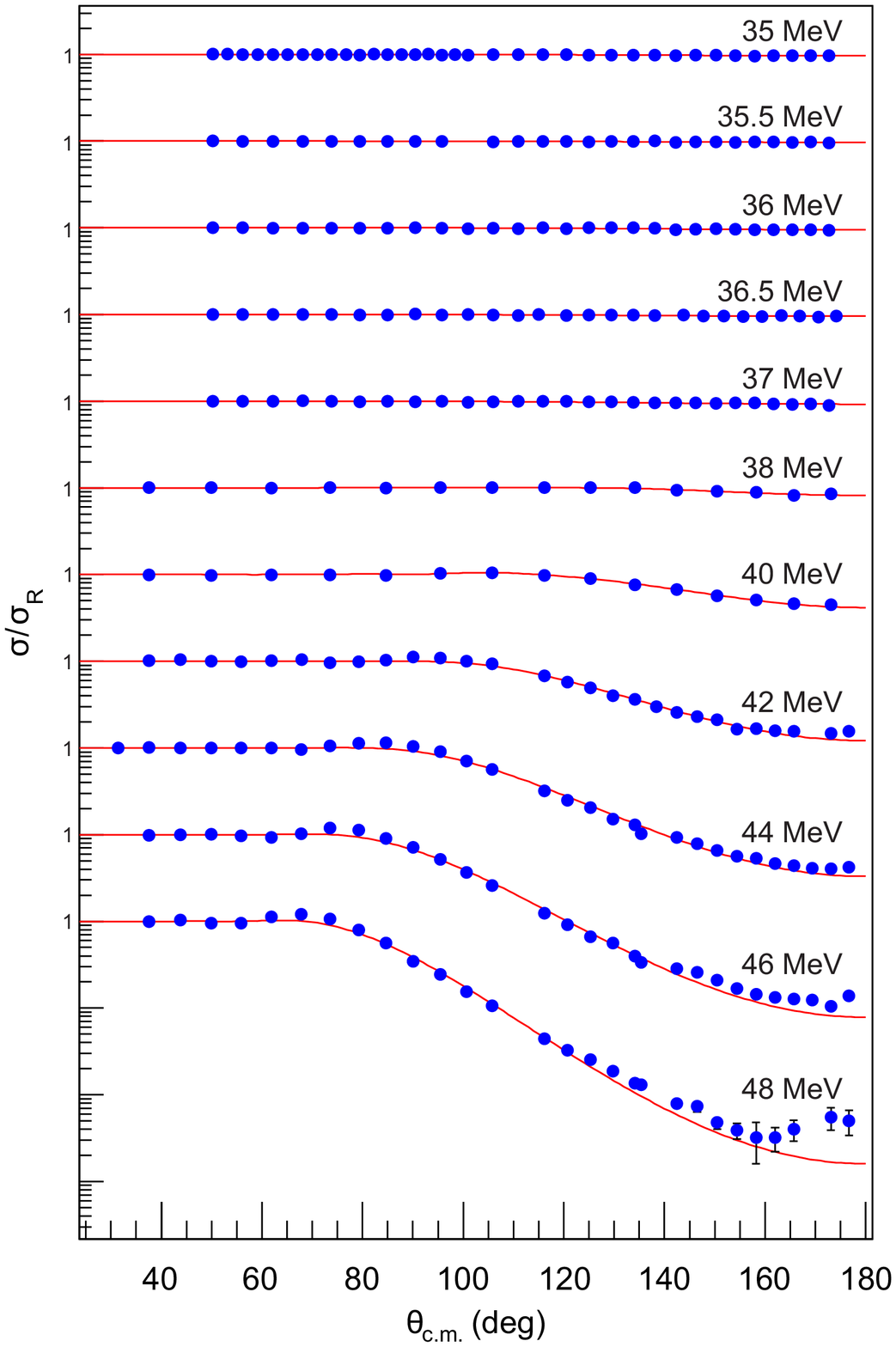}
\end{center}
\caption{\label{figNi} Angular distributions for the $^{16}$O+$^{58}$Ni system.}
\end{figure}

Eleven angular distributions from $35$ MeV to $48$  MeV have been analysed \cite{chamon2}. $V_{B}^{lab}=40.80$ MeV for this system.
The results are presented in Figure \ref{figNi} and in Table \ref{tabNi}. The small values and large errors of $V_0$ shown in Table
\ref{tabNi} for $E_{lab}=35-37$ MeV show that the potential is not determined at these energies below the Coulomb barrier.

\begin{table}[h]
\caption{\label{tabNi}Parameters of WS potential for the system $^{16}$O+$^{58}$Ni.}
\begin{center}
\begin{tabular}{c|c|c|c}
$E_{lab}$ (MeV) & $V_0$ (MeV) & $W_0$ (MeV) & $\chi^2$ \\
35.0 &	0.10$\pm$	4.35 &	3.84$\pm$	0.41 &	1.27 \\
35.5 &	0.10$\pm$	1.51 &	3.35$\pm$	0.27 &	1.51 \\
36.0 &	0.10$\pm$	3.67 &	3.54$\pm$	0.35 &	0.91 \\
36.5 &	2.56$\pm$	4.43 &	2.63$\pm$	0.80 &	2.01 \\
37.0 &	2.00$\pm$	0.06 &	3.48$\pm$	0.60 &	1.98 \\
38.0 &	15.95$\pm$     	0.79 &  1.00$\pm$	0.19 &	0.66 \\
40.0 &	17.59$\pm$	0.002 &	1.18$\pm$	0.09 &	1.06 \\
44.0 &	11.79$\pm$	0.0006 & 7.61$\pm$	0.004 &	9.48 \\
46.0 &	11.05$\pm$	0.007 &	 9.68$\pm$	0.009 &	13.17 \\
48.0 &	11.42$\pm$	0.0002 & 11.11$\pm$	0.0005 & 9.03 \\
\end{tabular}
\end{center}
\end{table}

\subsection{$^{9}$Be+$^{64}$Zn system}

\begin{figure}
\begin{center}
\includegraphics[width=18pc]{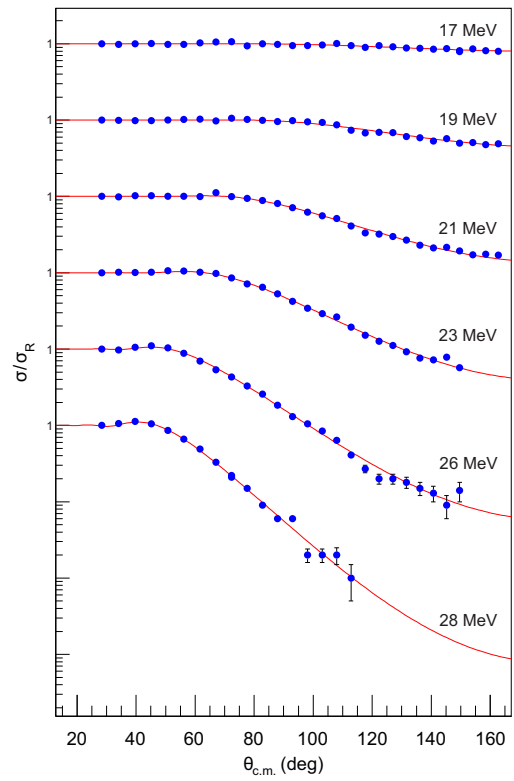}
\end{center}
\caption{\label{fig3} Cross section for the ${^9}$Be+$^{64}$Zn system. The energy values are shown in table \ref{tab2}. The experimental data is from Ref. \cite{moraes}. }
\end{figure}

Six angular distributions from $17$ MeV to $28$ MeV in the laboratory system have been analysed \cite{moraes}.
The Coulomb barrier is at $19.4$ MeV in the laboratory system (Eq. \ref{vcb}) and the results are presented in Figure \ref{fig3} 
and the resulting parameters are in table \ref{tab2}. 
 
\begin{table}[h]
\caption{\label{tab2}Parameters of WS potential for the system $^{9}$Be+$^{64}$Zn.}
\begin{center}
\begin{tabular}{c|c|c|c}
$E_{lab}$ (MeV) & $V_0$ (MeV) & $W_0$ (MeV) & $\chi^2$ \\
17.0 &	0.000 $\pm$	4.90 &  36.21$\pm$	3.17 &	1.07 \\
19.0 &	12.09$\pm$	1.43 &	19.17$\pm$	1.75 &	1.96 \\
21.0 &	12.06$\pm$	0.29 &	18.88$\pm$	0.61 &	4.22 \\
23.0 &	10.61$\pm$	0.16 &	14.51$\pm$	0.40 &	2.52 \\
26.0 &	10.12$\pm$	0.18 &	15.98$\pm$	0.79 &	1.62 \\
28.0 &	13.01$\pm$	0.25 &	12.47$\pm$	0.49 &	3.40 \\
\end{tabular}
\end{center}
\end{table}

\subsection{$^{9}$Be+$^{89}$Y system}

Seven angular distributions have been analysed for the $^9$Be$+^{89}$Y system. 
The laboratory energies range from $19$ MeV to $33$ MeV and $V_{B}^{lab}=23.25$ MeV. The results
are presented in Table \ref{tab3}. The values of the imaginary part of the potential drop down to energies lower than the Coulomb barrier, 
except for the lowest energy.
 
\begin{figure}
\begin{center}
\includegraphics[width=18pc]{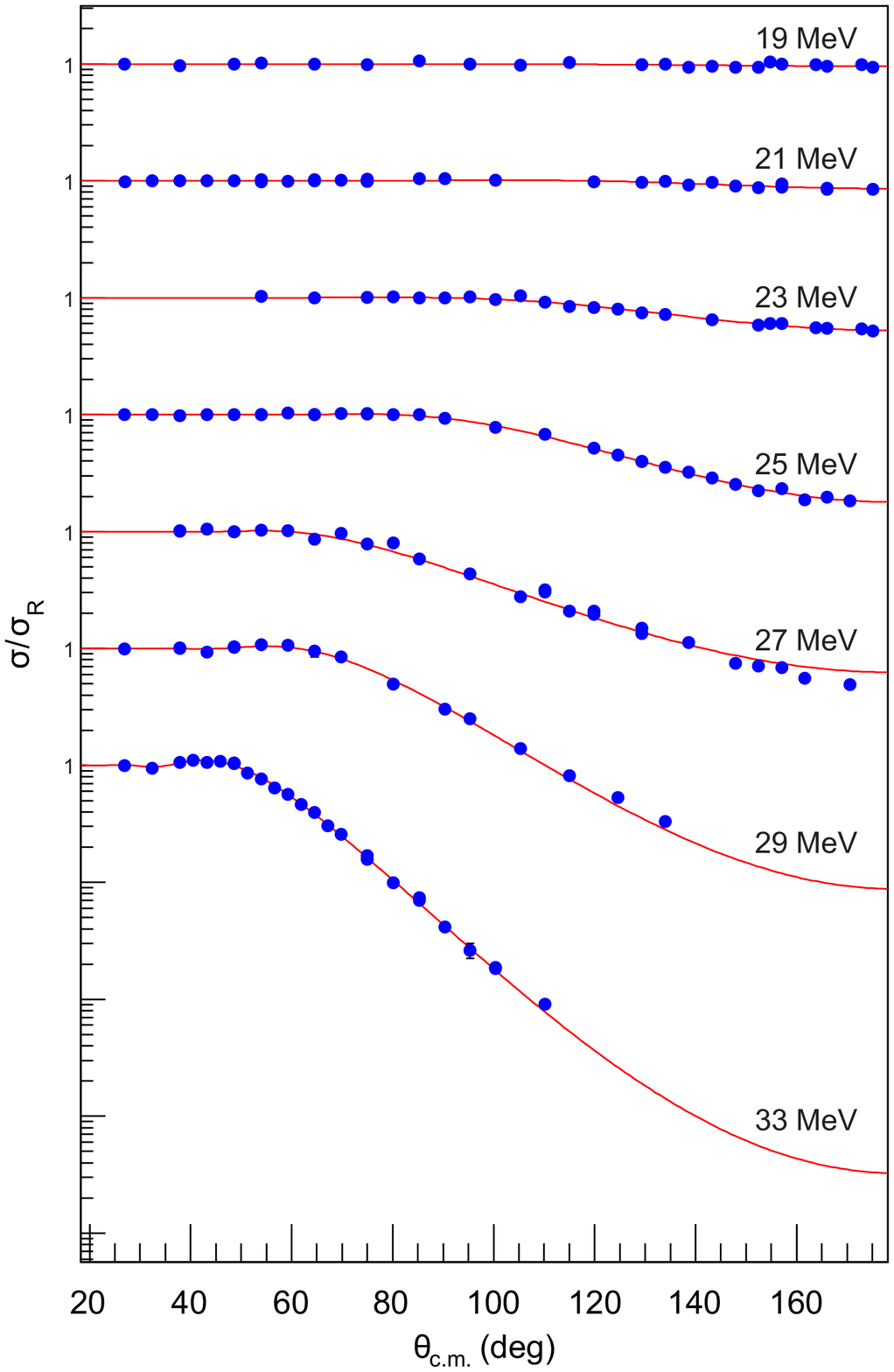}
\end{center}
\caption{\label{fig4} Cross section for the ${^9}$Be+$^{89}$Y system. The energy values are shown in table \ref{tab3}. The experimental data are from Ref. \cite{palshetkar}. }
\end{figure}

\begin{table}[h]
\caption{\label{tab3}Parameters of WS potential for the system $^{9}$Be+$^{89}$Y.}
\begin{center}
\begin{tabular}{c|c|c|c}
$E_{lab}$ (MeV) & $V_0$ (MeV) & $W_0$ (MeV) & $\chi^2$ \\
19.0 &	13.26$\pm$	0.40 &	17.37$\pm$	0.40 &	2.70 \\
21.0 &	20.51$\pm$	0.92 &	1.36$\pm$	0.87 &	1.87 \\
23.0 &	10.07$\pm$	0.68 &	7.48$\pm$	0.71 &	1.51 \\
25.0 &	8.50$\pm$	0.01 &	7.93$\pm$	0.01 &	2.24 \\
27.0 &	2.52$\pm$	0.34 &	29.12$\pm$	0.72 &	17.46 \\
29.0 &	8.63$\pm$	0.16 &	11.59$\pm$	0.33 &	11.81 \\
33.0 &	14.42$\pm$	0.10 &	15.11$\pm$	0.13 &	11.74 \\
\end{tabular}
\end{center}
\end{table}

\subsection{$^{9}$Be+$^{144}$Sm system}

\begin{figure}
\begin{center}
\includegraphics[width=18pc]{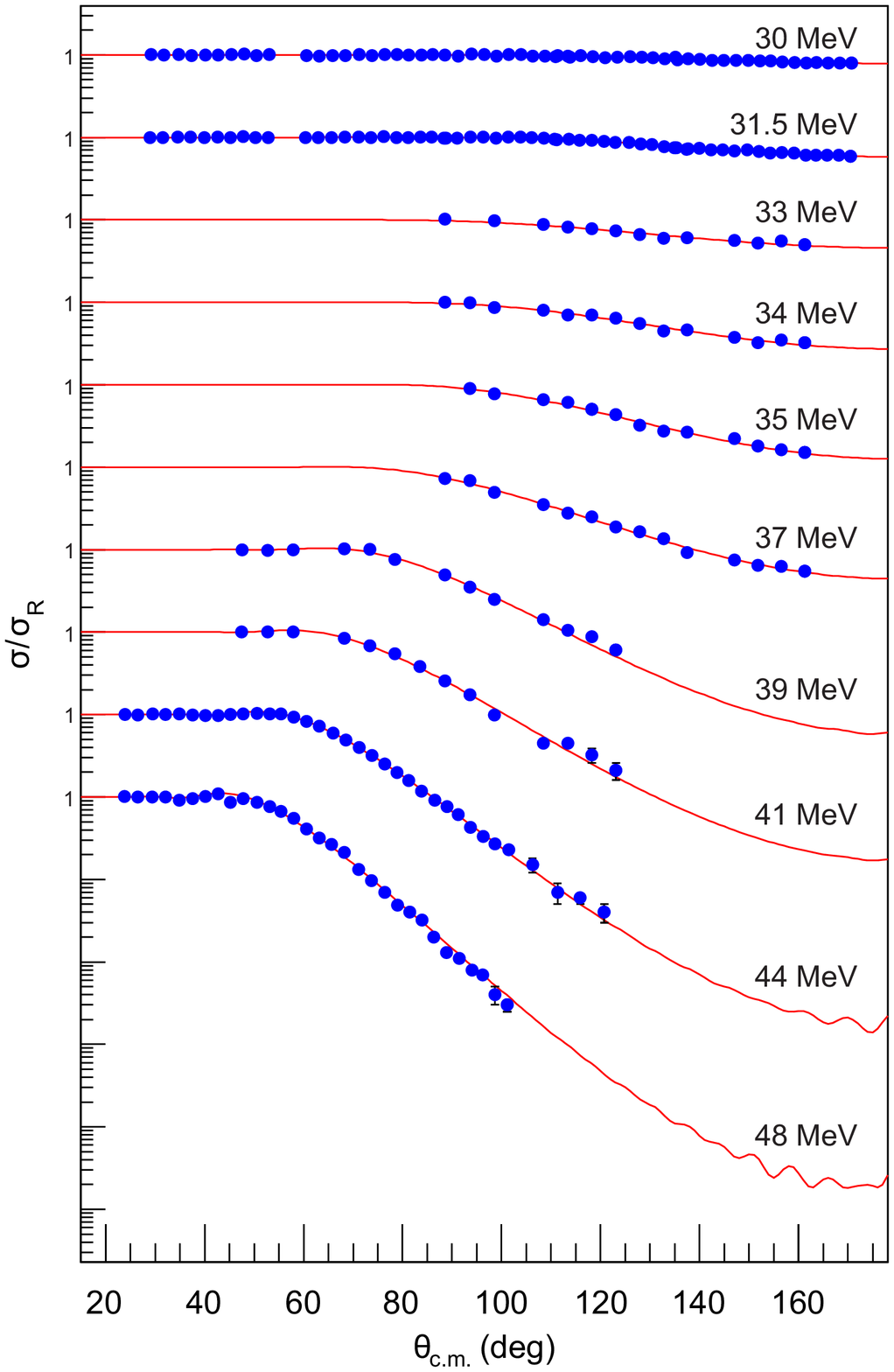}
\end{center}
\caption{\label{fig5} Cross section for the ${^9}$Be+$^{144}$Sm system. The energies values are shown in table \ref{tab4}. The experimental data are from Ref. \cite{gomes1,gomes}. }
\end{figure}

Ten angular distributions for the $^{9}$Be+$^{144}$Sm system have been analysed \cite{gomes1,gomes}. The energies range from below to above the Coulomb barrier at 
$V_{B}^{lab}=33.24$ MeV. See Figure \ref{fig5} and  Table \ref{tab4}. 

\begin{table}[h]
\caption{\label{tab4}Parameters of WS potential for the system $^{9}$Be+$^{144}$Sm.}
\begin{center}
\begin{tabular}{c|c|c|c} 
$E_{lab}$ (MeV) & $V_0$ (MeV) & $W_0$ (MeV) & $\chi^2$ \\
30.0 &	11.99$\pm$	2.55 &	15.28$\pm$	1.09 &	1.16 \\
31.5 &	13.94$\pm$	0.81 &	12.76$\pm$	0.57 &	1.32 \\
33.0 &	0.81$\pm$	4.24 &	16.76$\pm$	2.47 &	0.98 \\
34.0 &	7.44$\pm$	1.59 &	13.03$\pm$	1.62 &	0.73 \\
35.0 &	9.97$\pm$	0.83 &	13.41$\pm$	1.31 &	0.92 \\
37.0 &	7.42$\pm$	0.58 &	17.00$\pm$	1.11 &	0.96 \\
39.0 &	11.86$\pm$	0.42 &	13.41$\pm$	0.78 &	2.58 \\
41.0 &	11.63$\pm$	0.45 &	15.81$\pm$	1.04 &	1.87 \\
44.0 &	13.66$\pm$	0.18 &	15.67$\pm$	0.41 &	1.26 \\
48.0 &	13.83$\pm$	0.16 &	16.86$\pm$	0.35 &	14.34 \\
\end{tabular}
\end{center}
\end{table}

\subsection{$^{9}$Be+$^{208}$Pb system}

Fourteen angular distributions have been analysed in an energy around and below the Coulomb barrier \cite{yu}.
$V_{B}^{lab}=40.40$ MeV for this system. 
\begin{figure}[h]
\begin{center}
\includegraphics[width=18pc]{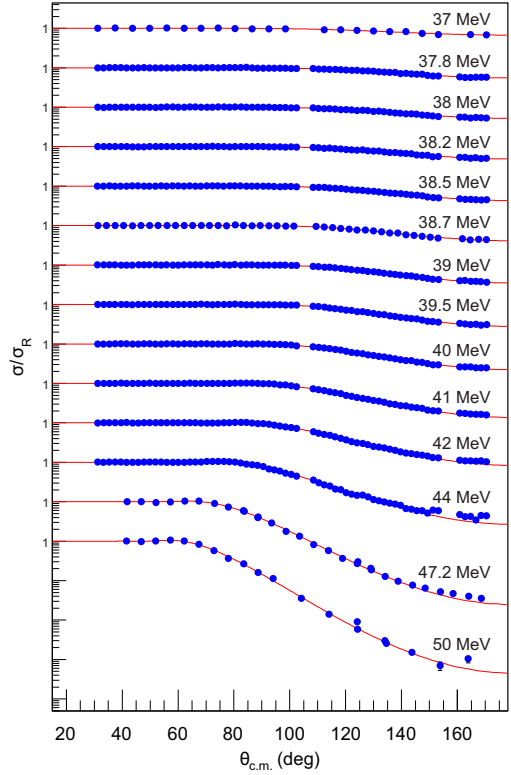}
\end{center}
\caption{\label{fig6} Cross section for the ${^9}$Be+$^{208}$Pb system. The parameters are shown in table \ref{tabPb} and the experimental data are from Ref. \cite{yu}. }
\end{figure}

\begin{table}[h]
\caption{\label{tabPb}Parameters of WS potential for the system $^{9}$Be+$^{208}$Pb.}
\begin{center}
\begin{tabular}{c|c|c|c} 
$E_{lab}$ (MeV) & $V_0$ (MeV) & $W_0$ (MeV) & $\chi^2$ \\
37.0 &	0.10 $\pm$ 2.77 & 29.38$\pm$ 0.56 & 6.34 \\
37.8 &	13.95$\pm$ 0.84 & 21.58$\pm$ 0.48 & 2.12 \\
38.0 &	15.98$\pm$ 0.64 & 19.98$\pm$ 0.39 & 3.37 \\
38.2 &	17.52$\pm$ 0.54 & 18.69$\pm$ 0.36 & 3.91 \\
38.5 &	18.39$\pm$ 0.37 & 18.53$\pm$ 0.28 & 3.59 \\
38.7 &	18.01$\pm$ 0.53 & 17.27$\pm$ 0.42 & 3.34 \\
39.0 &	19.08$\pm$ 0.25 & 17.18$\pm$ 0.22 & 7.60 \\
39.5 &	18.20$\pm$ 0.20 & 17.79$\pm$ 0.20 & 5.02 \\
40.0 &	16.68$\pm$ 0.18 & 19.06$\pm$ 0.20 & 7.07 \\
41.0 &	14.84$\pm$ 0.07 & 19.76$\pm$ 0.01 & 9.17 \\
42.0 &	13.01$\pm$ 0.10 & 20.69$\pm$ 0.14 & 18.00 \\
44.0 &	11.94$\pm$ 0.19 & 20.89$\pm$ 0.28 & 8.01 \\
47.2 &	14.50$\pm$ 0.07 & 19.84$\pm$ 0.18 & 13.60 \\
50.0 &	15.15$\pm$ 0.11 & 22.81$\pm$ 0.24 & 28.47 \\
\end{tabular}
\end{center}
\end{table}

\section{Fusion and total reaction cross section}

\begin{figure}
\begin{center}
\includegraphics[width=16pc]{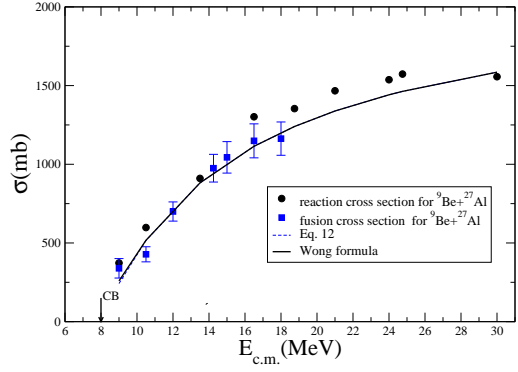}
\end{center}
\caption{\label{reac1} Comparison between the total reaction, fusion cross section with predictions of formula \ref{wong} and \ref{react} for the $^9$Be$+^{27\!}$Al system with  $R_B=8.29$ fm, $V_B^{(cm)}=7.99$ MeV and $\hbar w=3.25$ MeV. 
The fusion cross sections were taken from ref. \cite{marti} and the reaction cross section are from the optical model
calculations from section \ref{sec-27al}.}
\end{figure}

\begin{figure}
\begin{center}
\includegraphics[width=16pc]{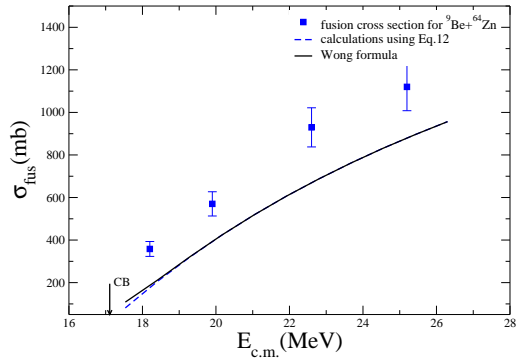}
\end{center}
\caption{\label{reac2}Comparison between the total  fusion cross section with predictions of formula \ref{wong} and \ref{react} for the $^9$Be$+^{64}$Zn system
with  $R_B=9.28$ fm, $V_B^{(cm)}=17.0$ MeV and $\hbar w=3.44$ MeV. 
The fusion cross sections are taken from ref. \cite{moraes}.}
\end{figure}

\begin{figure}
\begin{center}
\includegraphics[width=16pc]{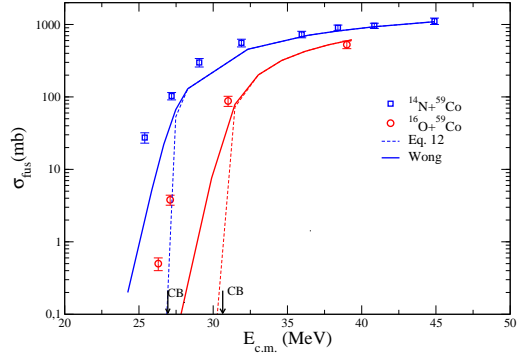}
\end{center}
\caption{\label{reac3}Comparison between the total fusion cross section with predictions of formula \ref{wong} and \ref{react} for the $^{14}$N$+^{59}$Co 
and  $^{16}$O$+^{59}$Co systems. The fusion cross sections are taken from ref. \cite{gomes1}.}
\end{figure}

\begin{figure}
\begin{center}
\includegraphics[width=16pc]{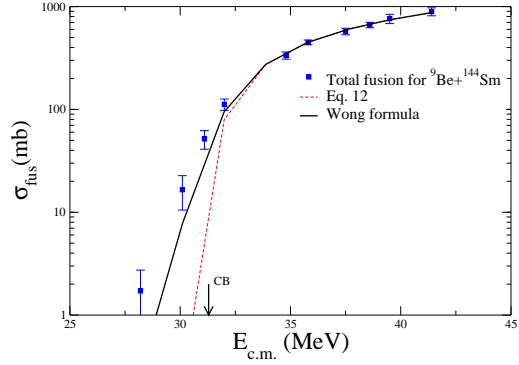}
\end{center}
\caption{\label{reac4}Comparison between the total  fusion cross section with predictions of formula \ref{wong} and \ref{react} for the $^{9}$Be$+^{144}$Sm 
system \cite{gomesx}. $R_B=10.67$ fm, $V_B^{(cm)}=31.29$ MeV and $\hbar w=3.54$ MeV. The fusion cross sections are taken from ref. \cite{gomesx}.}
\end{figure}

\begin{figure}
\begin{center}
\includegraphics[width=16pc]{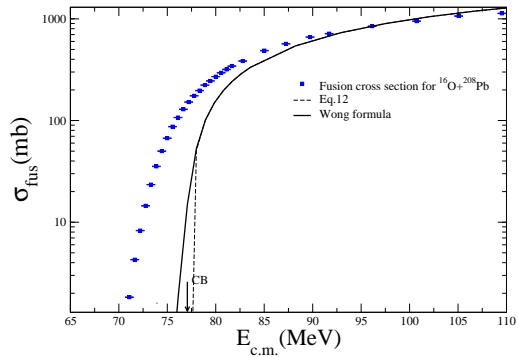}
\end{center}
\caption{\label{reac5}Comparison between $^{16}$O$+^{208}$Pb fusion cross sections \cite{dasgupta} and the calculations using the Wong formula and $R_B=11.68$ fm, $V_B^{(cm)}=77.07$ MeV and $\hbar w=2.45$ MeV.}
\end{figure}

The total reaction cross section is an important information that can be obtained from the elastic scattering. In addition, at energies around the Coulomb barrier, in many cases, 
fusion exhausts most of the total reaction cross section and can be estimated by barrier penetration calculations.
The well known Wong formula \cite{wong} for fusion has been applied with success to provide estimations of the fusion cross section.
\begin{equation}
\sigma_{wong}=\frac{R_B^2\hbar w}{2E_{c.m.}}\ln\{1+\exp[2\pi(E-V_B)/\hbar w]\} \label{wong}
\end{equation} 
This formula depends on 3 parameters, the Coulomb barrier position, height and its curvature ($\hbar w$),  the same parameters that have been determined 
on Sec.III. For energies above the Coulomb barrier the Wong formula reduces to a simpler one which depends only on two parameters, the position and height of 
the Coulomb barrier.
\begin{equation}
\sigma=\pi R_B^2(1-\frac{V_B}{E}) \label{react} 
\end{equation}

We applied formulas \ref{wong} and \ref{react} using the parameter calculated from formulas \ref{eq::rb1}, \ref{vcb}, \ref{hw} and \ref{2cnd}
for the $^9$Be$+^{27\!}$Al, $^9$Be$+^{64}$Zn  and $^{16}$O$+^{208}$Pb systems.
The results are shown in figures \ref{reac1}, \ref{reac2}, \ref{reac3}, \ref{reac4} and \ref{reac5}. 

We see that formulas  \ref{wong} (solid) and \ref{react} (dashed)  give the same result as the 
energy overcomes the Coulomb barrier. The agreement between the experimental fusion cross section and calculation is reasonable for energies above the Coulomb barrier.
For energies below the barrier the calculations predict a cross section much smaller than the experimental one as can be seen in Figure \ref{reac3}.
This is expected since it is well known that for energies below the Coulomb barrier the fusion is strongly affected by coupled channels effects, which are 
obviously not taken into account by the simple formulation presented here.

\section{Conclusions}

The real part of a Woods-Saxon potential that fits a double folding potential in the surface region ($r>R$) has been obtained. 
It was found that the tail of the double folding potential can be very well reproduced in all cases analysed here using a Woods-Saxon potential with a fixed geometry 
$r_0=1.3$ fm and $a=0.65$ fm and depths varying between $10-20$ MeV. There is a continuum ambiguity between $r_0$ and $V_0$.

A simple analytical formula has been derived using the real potential with depth $V_0=15$ MeV, which provides the position and height of the Coulomb barrier in very good agreement
with the double folding potential predictions. It is shown that the Coulomb barrier position and height depend on a single dimensionless parameter
$x$, which can be easily calculated as a function of the masses and charges of the colliding nuclei.

An optical model analysis has been performed for several systems using such potential, fixing the geometry for the real and imaginary parts, 
and adjusting the depths $V$ and $W$ to fit the angular distributions. It is shown that the potential proposed here provides reasonable fits 
of the scattering angular distributions for several stable systems at several energies above and below the Coulomb barrier. A strong variation
of $V_0$ and $W$ is observed at energies around the Coulomb barrier, as expected, due to the closing of the reaction channels as the energy 
goes down below the Coulomb barrier.

A criticism could be done to the optical model analysis presented here since, the obtained potentials, are not anymore strictly equivalent to the double folding, not even
in the suface region, because of the free variation of the real and imaginary depths. This is true and, it must be indeed just like that since the optical potentials that 
reproduce the data are not anymore the bare potential but the total optical potential, which includes all the polarization effects from the couplings
with other reaction channels.

Total reaction and fusion cross sections have been calculated using the analytical formula derived here and the result is compared with experimental 
fusion cross sections.
It  is  shown that, above the Coulomb barrier, the analytical formula provide a good approximation
for the fusion cross sections and the calculations can be done with a simple pocket calculator without the need of numerical computations.

\subsection*{Acknowledgements}
The authors wish to thank the Funda\c{c}\~ao de Amparo \`a Pesquisa do Estado de S\~ao Paulo (FAPESP) and the Conselho Nacional de Desenvolvimento 
Cient\'ifico e Tecnol\'ogico (CNPq), Coordena\c{c}\~ao de Aperfei\c{c}oamento de Pessoal de N\'{\i}vel Superior (CAPES) and
Comiss\~ao Nacional de Energia Nuclear (CNEN) for financial support.

\end{document}